\documentclass[preprint2]{emulateapj}
\usepackage{graphicx,amsmath,url}
\usepackage{bm}
\usepackage{multirow}
\graphicspath{{./fig/}{./png/}}

\newcommand{\EQ}{\begin{equation}}
\newcommand{\EN}{\end{equation}}
\newcommand{\EQA}{\begin{eqnarray}}
\newcommand{\ENA}{\end{eqnarray}}

\newcommand{\Eq}[1]{Equation~(\ref{#1})}
\newcommand{\Eqs}[2]{Equations~(\ref{#1}) and~(\ref{#2})}

\newcommand{\Sec}[1]{Section~\ref{#1}}

\newcommand{\Fig}[1]{Figure~\ref{#1}}

\newcommand{\Figs}[2]{Figures~\ref{#1} and \ref{#2}}

\newcommand{\Tab}[1]{Table~\ref{#1}}

{}
{}
{}

{}
{}
{}
{}
{}
{}
{}
{}
{}
{}
{}
{}
{}
{}
{}
{}
{}
{}
{}
{}

{}
{}
{}

{}

{}
{}

\newcommand{\UU}{\mbox{\boldmath $U$} {}}

\newcommand{\BB}{\mbox{\boldmath $B$} {}}

\newcommand{\JJ}{\mbox{\boldmath $J$} {}}

\newcommand{\nab}{\mbox{\boldmath $\nabla$} {}}

\newcommand{\deriv}[3]{\frac{#3\hspace*{-.06em} {#1}}{#3\hspace*{.06em} {#2}}}

\newcommand{\parder}[2]{\deriv{#1}{#2}{\partial}}

\renewcommand{\v}[1]{{\boldsymbol #1}} 
\newcommand{\rot}{\v\nabla \times}
\newcommand{\parr}[2]{\frac{\partial #1}{\partial #2}}
\newcommand{\parrline}[2]{\partial #1/ \partial #2}
\newcommand{\parrx}{\partial_x}

\begin{document}

\title{Short Circuits in Thermally Ionized Plasmas: A Mechanism for
  Intermittent Heating of Protoplanetary Disks}
\author{Alexander Hubbard, Colin P. McNally\altaffilmark{1,2}, \& Mordecai-Mark Mac Low\altaffilmark{1}}
\affil{Department of Astrophysics, American Museum of Natural History, 79th St.\ at Central Park West, New York, NY 10024-5192, USA}
\email{ahubbard@amnh.org, cmcnally@nbi.dk, mordecai@amnh.org}
\altaffiltext{1}{Department of Astronomy, Columbia University, New York, NY, USA}
\altaffiltext{2}{Niels Bohr International Academy, Niels Bohr
  Institute, University of Copenhagen, Copenhagen, Denmark} 
\begin{abstract}
Many astrophysical systems of interest, including
protoplanetary accretion disks, are made of turbulent magnetized gas 
    with near solar metallicity. 
    Thermal ionization of alkali metals in such gas exceeds
    non-thermal ionization when temperatures climb above roughly
    1000~K. As a result,
the conductivity, proportional to the ionization fraction, gains 
    a strong, positive dependence
on temperature.
In this paper, we
     demonstrate that this relation between the temperature and the conductivity
     triggers an exponential instability that acts similarly to
an electrical short,
where the increased conductivity concentrates the current and locally increases the Ohmic heating.
This 
    contrasts with the resistivity increase expected in an ideal
    magnetic reconnection region.
The 
    instability
acts to focus narrow current sheets into even narrower
     sheets with far higher currents and temparatures.
We lay out the basic principles of this behavior
in this paper using protoplanetary disks as our example host system, motivated by 
observations of chondritic meteorites and their ancestors, dust grains in protoplanetary
disks, that
reveal the existence of strong, frequent heating events
    that this instability could explain.
\end{abstract}


\section{Introduction}

In this paper, we describe an exponential instability that acts to
narrow and strengthen current sheets in partially ionized gas with a
resistivity inversely dependent on temperature.
While the physics that we 
 describe pertains
to any magnetized system with both current sheets and
a conductivity that increases 
strongly enough
with temperature,
this problem arises specifically in the context of
protoplanetary disks, where the formation of high-temperature
minerals such as chondrules and crystalline silicates
suggests strong, intermittent heating events.

As the local temperature in such a disk climbs above
$\sim 1000$~K, the dominant source of free electrons
becomes the thermal ionization of
alkaline metals.  These
temperatures are still well below the ionization energy,
so the argument of the exponential in the Boltzmann term is quite small;
thus the ionization fraction $x_e=n_e/n_n$ depends steeply on
temperature. 
We have found that this results in startling new
behavior with positive temperature fluctuations
increasing conductivity, concentrating current sheets, and positively feeding back on
the temperature through enhanced
local Ohmic heating.  

This is quite different from classical reconnection, with electrical
short circuits effectively forming in these regions.
This effect is the opposite of the more commonly assumed anomalous
resistivity that increases in the reconnection region
\citep{KrallLiewer71,SatoHayashi79,Yamada10}.
(Classical reconnection was applied to disks using general
energetic arguments by \citealt{King10}.) 
While our mechanism narrows current sheets, similarly to ambipolar diffusion \citep{BZ94}, 
the mechanism is Ohmic resistivity rather than a drift
of the charge carriers relative to the neutral gas.
Our mechanism is also different from previous work on partially ionized reconnection because
that focused on the transition to the collisionless regime \citep{Malyshkin11, Zweibel11}.

Observations of protostellar disks have revealed their integrated
properties, including masses \citep{bs93} and accretion rates
\citep{h98}.   Compositional gradients in the dust \citep{vb04} can be
detected, as well as
the difference between predominantly amorphous and crystalline mineral
structures at different radii \citep{Waelkens96,Malfait98}. 
The observed presence of crystalline minerals at large
 radii \citep{Sargent09} suggests the need for a heating mechanism active in
 the disk far from the parent star.
Meanwhile,
we have direct evidence of conditions in the protosolar disk.  Laboratory measurements of textural, mineralogical, chemical,
and isotopic properties of meteoritic chondrules and calcium-aluminum rich
inclusions, 
as well as related high-temperature
materials in comet samples \citep{Brownlee06,Zolensky06,Nakamura08,Simon08}, give strong
constraints on their local formation history and environment. They
represent melts condensed and cooled from temperatures of 1500--1800~K at rates
of around 100--1000~K/hour
\citep{RadomskyHewins90,LofgrenLanier90,Connolly98,sk05,Ebel06}: far faster
than disk dynamical timescales, but far slower than the free-space
cooling time of millimeter sized objects. 
The source of these processed materials in unclear. While there is a vast reservoir of gravitational potential
energy in the disk, tapping it through an accretion flow
 to create hot regions of finite size, as appears to be needed, is non-trivial.

The primary source of angular momentum transport
in protoplanetary disks appears to be
the conversion of orbital kinetic energy into magnetic
energy through the magnetorotational instability
\citep[MRI;][]{Velikhov59,Chandrasekhar61,bh91,bh98}, 
with gravitational instability playing a larger role at earlier times
\citep[e.g][]{Lodato04}.  The ionization structure of the disk
midplane may introduce a magnetically dead zone
 at some radii and periods of disk history 
\citep{Gammie96}, with reduced though
non-zero turbulent viscosity \citep{Fleming03,Oishi09}. Its exact structure and history
remains controversial \citep{Glassgold97,Sano00,Ilgner06,Umebayashi09,Turner09}.

MRI draws on the huge amount of energy contained in the differential
rotation of the disk to drive magnetohydrodynamical (MHD) turbulence.
The turbulence will dissipate that energy into heat.  However, the
heating will occur intermittently, not uniformly.  MHD turbulence
forms current sheets \citep{Parker72,Parker94,1997PhR...283..227C}
that dissipate energy at far greater than the average
rate, and can provide locations for magnetic reconnection to occur.
\citet{Romanova11a} noted in passing the volume filling nature of this
mechanism \citep[also see][]{Romanova11}. \citet{Hirose11} used a
moderate resolution simulation to demonstrate that such current
sheets forming in the atmosphere above a dead zone can locally heat
gas well above the radiative equilibrium temperature. 
Those calculations, along with preliminary work of our own
\citet{McNally12a,McNally12b} suggests that such current sheets may
well heat the gas up to temperatures 
sufficiently high for the instability described here to set in.  

This is hardly the first suggestion that electromagnetic fields can drastically alter
the temperature profile in a protoplanetary 
disk. \citet{Levy89} examined reconnection heating in disks as a
chondrule formation mechanism. However, they worked
before the nature of the turbulence driving angular momentum
transport was understood. Therefore, they reasoned by analogy to
the Sun that a stratified, convective, magnetized flow would drive
reconnection in the low density region above it. Thus, they only
considered coronal heating many scale heights above the surface
of the disk. However, we now understand that the turbulence in
the disk is probably not convective but rather driven by the
differential rotation acting through the field, so that
intermittent dissipation will occur throughout the disk. 
The short-circuit instability is in 
many ways similar to lightning 
\citep{Whipple66,Horanyi95,Pilipp98,DC2000,Muranushi10,Muranushi12}: a rapid local increase in 
the ionization fraction leads to high currents and a dramatic release of energy.
However, there are significant differences in that the increase in ionization is thermal rather
than due to electric fields strong enough to directly induce ionization breakdown. Further, the dynamo magnetic fields
act as a current source rather than a voltage source. Electrical-short like
behavior is only possible due to the residual current that flows in the low
temperature regions around the instability, maintaining a non-trivial Ohmic electric field
(see \Sec{short:statement}). This allows us to bypass the need to generate.
electric fields strong enough to directly ionize the gas, which is
a non-trivial challenge to lightning models.
Another related proposed mechanism is melting of charged dust
by acceleration through standard reconnection regions
\citep{Lazerson10}.

In this paper,
we lay out the basic principles of this novel behavior, and explore
the implications for heating in dusty protoplanetary disks in more
detail in a companion paper \citet[][hereafter Paper II]{McNally12b}.  
Closely related behavior
in planetary atmospheres 
was found by \cite{Menou12},  who called it 
the thermo-resistive instability; 
though
that term was also used by
\cite{Price12} for a system where the resistivity increases with temperature.

In Sect.~\ref{sec:temp} we describe the physical principles at
play. In Sect.~\ref{sec:eqns} we lay out a numerical approach to
modeling this behavior, whose results are described in
Sect.~\ref{sec:results}, and discussed in Sect.~\ref{sec:discuss}.

\section{Temperature dependent resistivity}
\label{sec:temp}
\subsection{Spatially-varying resistivity}
Consider the induction equation in the presence of Ohmic resistivity $\eta$
\EQ
\parder{\BB}{t} = \rot \left(\UU \times \BB - \eta \JJ\right),
\EN
where $\BB$ is the magnetic flux, $\JJ$ the current, $\eta$ the
resistivity, and $t$ time.
If the resistivity is spatially uniform as normally assumed, then the resistive term can be rewritten:
\EQ
\parder{\BB}{t} = \rot \left(\UU \times \BB\right) + \eta \nab^2 \BB,
\EN
where the effect of the resistivity is to diffuse the magnetic field.
However, if the resistivity varies spatially, we must consider its spatial derivative:
\EQ
\parder{\BB}{t} = \rot \left(\UU \times \BB\right) + \eta \nab^2 \BB-\left(\nab \eta\right) \times \JJ.
\label{induction}
\EN
If the resistivity shows strong spatial variation, the final term in \Eq{induction} can
dominate over the diffusive one.  In the limit of a one-dimensional system, varying along $\v x$
with the magnetic field pointing along $\v y$, we have
\EQ
\parr{B_y}{t} = -\parrx \left[v_x B_y\right] + \eta \parrx^2 B_y  + \parrx \eta \parrx B_y . \label{induction2}
\EN
If the $\parrx \eta \parrx B_y$ term is dominant over the
diffusive term $\eta \parrx^2 B_y$, the Ohmic resistivity can act to steepen, rather than broaden,
magnetic field gradients.

This consideration of the spatial variation of the resistivity comes into full focus if the resistivity drops steeply
with temperature, which can occur in a mostly neutral plasma in
temperature ranges where one or more species is being thermally ionized.  In this case, a local positive
temperature perturbation that increases ionization will drive a
positive current perturbation, which can in turn result in
an increase in the local Ohmic heating.

\subsection{Steady State}
\label{short:statement}

The base problem is one common to elementary electrical circuits: electrical shorts, albeit in the
current driven regime.  To put this on a quantitative basis, let us
begin by considering a base steady-state
one-dimensional system of length $L$,
aligned with the $x$-axis and 
centered at $x=0$, with magnetic structure given by
\begin{align}
&B_y|_{\pm L/2}=\pm B_0/2 \label{B1D} \\
&J_z=\parrline{B_y}{x} \label{J1D} \\
&\parrline{B_y}{t} =\parrx \left(\eta J_z\right)=0. \label{dBdt1D}
\end{align}
Under the assumption of uniform $\eta=\eta_0$ we have simply
$J_z=B_0/L\equiv J_0$.  Note that we have assumed that the velocity is $u_x=0$.

We now perturb the system, decreasing the resistivity to $\eta=\eta'=\overline{\eta}\, \eta_0$
in a region of width $\Delta L=\overline{\delta} L$, centered at $x=0$.   Equations (\ref{B1D}) -- (\ref{dBdt1D})
then require
\begin{align}
&\left(1-\overline{\delta}\right) J_1+ \overline{\delta} J_1'= J_0 \\
&J_1 = \overline \eta J_1', 
\end{align}
from which we can derive
\EQ
J_1'=\left[\overline{\delta}+(1-\overline{\delta})\overline{\eta}\right]^{-1} J_0,
\EN
where $J_1$ is the value of $J_z$ in the region with $\eta=\eta_0$ and $J_1'$ is the value
of $J_z$ in the region with $\eta=\eta'$.  As long as $\overline{\eta}< 1$,
$J_1'>J_0$, which is quite natural as current will preferentially flow
in regions of low resistivity.

More interesting is the Ohmic dissipation.
The dissipation in the perturbed region  $\eta' J_1'^2$ always exceeds
that in the unperturbed region, as the electric fields in the two
regions are equal by the requirement of a steady state. It also
exceeds the dissipation in the base state $\eta_0 J_0^2$ if 
\EQ
\overline{\eta}^{1/2} > \frac{\overline{\delta}}{1-\overline{\delta}}. \label{needwings}
\EN
This implies that even small decreases in the resistivity result in increased heating
(effectively an electrical short) as long as $\overline{\delta} < 1/2$.  This
size limitation is required to maintain
sufficient residual current in the non-perturbed region that the electric field resistively generated
there can be the effective voltage source for the short.

In a partially
ionized medium where the resistivity decreases with temperature
because thermal ionization produces increased
charge carrier density,
one would expect the perturbed region to continue heating, reducing
its resistivity further.  Interestingly, this effect actually will
reduce the total energy dissipated:
the total current is constant, and we are reducing the resistivity of the medium the current
flows through.  While this effect generates local hot spots, it also decreases the total
rate at which magnetic energy is dissipated into heat.

\subsection{Instability Analysis}
\label{linearanalysis}

\Sec{short:statement} gives a qualitative picture of a system that
seems likely to experience an instability
that would lead to an ever narrowing current sheet with increasing temperature
and local heating.  A real system will not be so idealized of course.  We can explore the
instability conditions quantitatively by performing a linear stability
analysis on a slightly less
simplified one-dimensional model.  We assume an incompressible fluid
that cools (presumably radiatively) to a background bath temperature
$T_b$ in a time $t_b$.  The equations
are then
\begin{align}
&\parr{B_y}{t}=-\parrx \left[-\eta \parrx B_y\right], \\
&\parr{T}{t}=-\frac{T-T_b}{t_b}+\frac{T_0}{t_h} \frac{\eta \left(\parrx B_y\right)^2}
{\eta_0 J_0^2},
\end{align}
where the Ohmic
heating time is $t_h$.  The Ohmic heating term is normalized at the
reference temperature $T_0$ by the Ohmic heating with $\eta=\eta_0$
and $J=J_0$.  More exactly, we define the heating time
\EQ
t_h=\frac{4 \pi n_n k_B T_0}{\left(\gamma-1\right) \eta_0 J_0^2}, \label{theat}
\EN
where the factor of $4 \pi$ comes from our use of cgs electromagnetic units,
$n_n$ is the neutral number density and we assume a low ionization fraction.
We track the temperature dependence of the resistivity through the 
general equation
\EQ
\left.\parr{\eta}{T}\right|_{T_0}= -\frac{\eta_0}{T_1}, \label{etatempdep0}
\EN
where $T_1$ parameterizes the strength of the temperature gradient of $\eta$.
This also allows us to define the heating time of the resistivity $t_{\eta'}$ through
\EQ
t_{\eta'} \equiv \eta_0 \left|\frac{T_0}{t_h} \left.\parr{\eta}{T}\right|_{T_0}\right|^{-1} = \frac{T_1}{T_0} t_h.
\label{etatempdep1}
\EN
In a dominantly neutral medium in LTE, the Saha equation is a simple approximation to the thermal ionization behavior,
and the resistivity is dominated by the ionization fraction (see Equations \ref{spitzerresistivity} and \ref{ionfrac} for more detail).
In that case, considering only the exponential term in Equation~\ref{ionfrac} for analytic simplicity, 
\Eq{etatempdep0} becomes
\EQ
\left. \parr{\eta}{T}\right|_{T_0} \simeq \left(-\frac{T_i}{T_0^2}\right) \eta_0,
\label{etatempdep2}
\EN
where $T_i$ is the temperature associated with first ionization .
With $T_i=2.5188 \times 10^4$~K, the ionization temperature of potassium, this approximates
 the ionization behavior of protoplanetary disks at $T\sim
1000$~K, as potassium has a low
first ionization energy ($k_B T_i$) and sufficient abundances \citep{Fromang02}.
In this case, \Eq{etatempdep1} becomes
\EQ
t_{\eta'}=\frac{T_0}{T_i} t_h.
\label{etatempdep3}
\EN

With these approximations and definitions, we can derive the dispersion relation for a perturbation of the form
$e^{ikx+\lambda t}$ applied to  a base state with
\begin{align}
& B_y(x)=J_0 x \\
& T(x)=T_0.
\end{align}
The linearized equations for the perturbations are
\begin{align}
&\lambda \triangle B =\frac{\eta_0 J_0 i k \triangle T}{T_1} - \eta_0 k^2 \triangle B \label{linB}\\
&\lambda \triangle T =-\frac{\triangle T}{t_b}-\frac{2 \triangle T}{t_{\eta'}}+2 \frac{T_0}{t_h} \frac{ik\triangle B}{J_0}. \label{linT}
\end{align}
Solving \Eqs{linB}{linT}
we find
\EQ
\lambda^2+\lambda \left[\frac{1}{t_r}+\frac{1}{t_b}+\frac{1}{t_{\eta'}}\right]+
\frac{1}{t_r}\left[\frac{1}{t_b}-\frac{1}{t_{\eta'}}\right]=0, \label{instabgrowth}
\EN
where $t_r \equiv k^2/\eta_0$ is the resistive time of the
perturbation. 
As $t_r, t_{\eta'}$ and $t_b$ are all positive, it follows
that the perturbation can exhibit exponential growth ($\lambda>0$) if the constant term in \Eq{instabgrowth} is negative,
i.e. if
\EQ
t_{\eta'}< t_b, \label{instabcriterion}
\EN
which condition is independent of $t_r$,
unlike the usual situation for reconnection.  
While in the following section we will assume
a resistivity profile that gives \Eq{etatempdep2}, the existence of the instability requires only
a strong enough gradient of the temperature dependence of $\eta$ (as measured through $T_1$
and $t_{\eta'}$).

The independence of \Eq{instabcriterion} on $t_r$ 
arises in part because the resistivity plays an equal role in in the resistive time $t_r$ and the Ohmic
heating time $t_h$ and in part because the magnetic field transport is mediated purely through
resistive effects in the imposed absence of velocity.
The steepening of magnetic field gradients through this instability
and standard resistive spreading of magnetic fields occur through the
same resistivity operator. 
We emphasize this point: the transport of magnetic fields into the dissipation region
is resistive in nature, rather than advective.  Accordingly, this mechanism does
not immediately struggle with the problem of exhaust
that has bedeviled attempts to understand observed fast
reconnection in the solar corona and elsewhere.

The above analysis is a significant simplification of actual physical
systems even beyond its one-dimensional nature.  In particular, we
note that in a physical reconnection region, the background current is
not constant in time, and would be expected to decay resistively until
the unstable modes have had time to grow.
This occurs through the diffusive term in \Eq{induction2} applied to the background current,
which have been set to $0$ by our choice of background state, but which will not be negligible in
general.

As we will see, the assumption of 
a single physical length scale imposed in the above analysis also 
breaks down in practice, with the high temperature region narrowing
over time in the non-linear regime.
Further, MRI-active
protoplanetary disks are compressible and expected to have minimal plasma
$\beta\sim 1$--$10$.  The simulations described in the next section
include these
complications by implementing terms such as the Lorentz force, which acts to compress a
current sheet, and adiabatic heating and cooling.  
Finally, we note that our analysis of the current sheet has been done 
along the shortest dimension, and 
that the current sheet will be much larger in the perpendicular, neglected dimensions.
Because of this, the
approximation of cooling to a bath temperature is a noteworthy
oversimplification, and any cooling may act to expand the high
temperature regions by radiative heating of the surroundings, 
as treated in more detail in Paper~II.

\section{Equations and numerical methods}
\label{sec:eqns}
\subsection{Numerical methods}

To model the dramatic behavior suggested by the linear analysis, we
have written a one dimensional code using sixth order finite differences
 on a logarithmic grid.  We use implicit time integration with the 
CVODE package \citep{Hindmarsh05}.
This allows us to follow the large, spatially
limited variations in the resistivity. 
The logarithmically
spaced grid allows us to push the boundaries towards infinity while retaining
resolution in the center of the current sheet.  The number of grid
points used in the simulations reported here varies from 500
to 1000. Only the right half of the domain is included ($x>0$) and a symmetrical inner 
boundary condition is used.

\subsection{Equations}

We solve the MHD fluid equations in 1.5 dimensions, including
$x$-gradients and $y$-components of vectors,
with the one non-ideal term
being the Ohmic resistivity.  We use a somewhat more exact model of
thermal ionization dominated by potassium. 
Although the linear analysis presented in \Sec{linearanalysis} considered cooling,
for this model we neglect cooling terms, deferring that
additional complexity to
Paper~II. 
We use cgs units: magnetic field in Gauss and density in $\mathrm{g}/\mathrm{cm}^{-3}$.

With these approximations, the MHD equations become
\begin{align}
&\parr{\rho}{t} = -\parrx \left(\rho v_x\right), \\
&\parr{v_x}{t}  = -v_x \parrx v_x-\frac{1}{\rho} \parrx P -\frac{1}{8\pi \rho} \parrx B_y^2
+\parrx \zeta_S \parrx v_x, \\
&\parr{B_y}{t} = -\parrx \left[v_x B_y - 
\eta(x) \parrx B_y\right], \\
&\parr{T}{t}      = -\parrx \left(T v_x\right) - c_TP v_x \nonumber \\
&\quad \quad \quad \quad  +\frac {c_T \eta}{4\pi \rho} \left(\parrx B_y\right)^2+c_T \zeta_S \left(\parrx v_x\right)^2,
\end{align}
where $\zeta_S$ is a shock viscosity included for stability.  
The shock viscosity $\zeta_s$ is given by
\begin{align}
\zeta_s = C_s \max(-\partial_x v_x \Delta x^2)_+  
\end{align}
where the constant $C_s$ is taken as $10$, $\max()_+$ denotes taking the maximum positive value over five grid points or zero otherwise, and $\Delta x$ is the grid spacing.
The equation of state
is that of an ideal gas and $c_T$ is the conversion factor between temperature
and energy:
\begin{align}
&P = n_n k_B T, \\
&c_T \equiv \frac{\left(\gamma-1\right) m_n}{k_B},
\end{align}
where we use
$\gamma=7/5$, $n_n$ is the neutral number density (assumed to dominate) and $m_n$ is the 
neutral molecular mass.  The resistivity associated with a dominantly neutral gas
is given by \cite{Balbus01}
\EQ
\eta = 234 T^{1/2} x_e^{-1} \rm{cm}^2/\rm{s} \label{spitzerresistivity}
\EN
and the ionization fraction $x_e \equiv n_e/n_n$, under the assumption  
that the species being ionized is predominantly neutral and thermally ionized, becomes
\begin{align}
x_e = 8.7 \times 10^9~a^{1/2} \left(\frac{n_n}{1~\mathrm{cm}^{-3}}\right)^{-1/2}&\left(\frac{T}{10^3\rm{K}}\right)^{3/4} \nonumber \\
&\times \exp\left(-\frac{T_i}{T}\right). \label{ionfrac}
\end{align}
where $a$ is the fraction of the ionizing species to the total neutral population.
In our canonical model, we consider only the thermal ionization of potassium \citep{Fromang02}.

At the densities $\rho \sim 10^{-9}$g/cm$^{3}$, mean molecular mass $\mu=2.33$ and
potassium fraction $a=10^{-7}$ of our canonical model, this equation breaks down
at $T\gtrsim 1600\,$K when the potassium is significantly ionized.  At higher temperatures
in protoplanetary disks,
other metals will also begin to contribute to the free electrons.
At lower temperatures, below $T\sim 1000\,$K, the ionization fraction
from thermal processes becomes
so low that in any astrophysical system some non-thermal ionization source,
such as ionizing stellar radiation or radionuclide decay,
will dominate over thermal ionization.  Even if they did not, the physical length scales required
to achieve any MHD action in the presence of so high a resistivity become
absurd.
While for physical purposes, \Eq{ionfrac} only applies in the
temperature range $1000\,$K--$1600\,$K, 
we will consider evolution at starting temperatures of $500\,$K and
$2000\,$K to help test predictions about the strength of the gradient of the resistivity with respect to temperature.

\subsubsection{Initial and boundary conditions}

We consider initial conditions with $v_x=0$,
$\rho_0 = 10^{-9}$~g~cm$^{-3}$, $a = 10^{-7}$
and
\EQ
B_y(x)=B_0' \tanh(x/\ell_0), \label{initialB}
\EN
which reproduces the magnetic field of a \citet{Harris62} current sheet.
The density and potassium abundances are inspired by
conditions in the midplanes of protoplanetary disks with
active MRI
\citep{Boss96,Sano00}.
We denote the initial conditions at the box edge with the subscript $0$,
and use box widths large enough compared to $\ell_0$ that $B_0$ and $B_0'$
are functionally identical.
The density and temperature are then set to counterbalance the Lorentz
force in the center assuming an adiabatic compression.
$B_0$ is a control parameter that sets the total magnetic energy in the simulation.
As we are interested in the 
ability of the magnetic field to heat
the gas, instead of labeling runs with $B_0$, we label them with the
plasma beta: the ratio of thermal to magnetic pressure
\EQ
 \beta_0=8\pi \rho_0 k_B T_0/(\mu m_p B_0^2).
\EN
A value of $\beta_0 \sim 1$ signifies an initial magnetic field energy that could
raise the temperature throughout the box by $\sim50\%$ if converted directly to heat.
The conversion  is, however,
localized in our models resulting in intermittent regions with substantially higher temperatures.

We list our control parameters in \Tab{runpars}.
A characteristic driving scale
for turbulence in the inner portion of a 
protoplanetary disk would be
$L \sim 5 \times 10^{10}$~cm, estimated at $1$~AU for a Shakura-Sunyaev
$\alpha=10^{-2}$ \citep{SS73}.
The accretion luminosity and the irradiation from the central star
are inadequate to raise the volume averaged temperatures to
$T \sim 1000$~K at that position for prolonged periods.  Still, temperature spikes,
through magnetic reconnection, shocks or accretion events could all
contribute to the temperature structure, leaving $1$~AU a convenient
scale: just far enough from the central star that
temperature excursions
are unlikely to hit the $1000$~K threshold.
Observations of protoplanetary disks 
   also 
show an inner wall where dust is sublimated near
$1500$~K, so the existence of regions where the temperature 
   exceeds
$1000$~K is not in doubt
\citep[e.g.][]{Dullemond10}.

The resistive time scales that we derive drop quickly with
increasing temperature up to 1500~K.  This leads to the concern
that resistive diffusion might destroy current sheets before the
ionization instability can grow.  However, the
growth time of the instability depends linearly on the resistive
time (see \Eq{instabgrowth2}), so that as the
resistive time drops, the instability grows proportionally
faster.  The growth time,  
when measured in terms of the resistive time, does increase  
with the plasma $\beta$ 
(see discussion in Sect.~\ref{sec:growth}, where $t_c$ is a
growth time estimate). 
Values of the plasma $\beta$ of the order of 3--10 have been
reported to occur above the midplane in moderate  resolution models
of MRI-active disks (\citealt{Flock11}, Figure 11). 
The main constraint on instability (given reasonable protoplanetary values of $\beta$),
is that the background system vary slowly
compared to the resistive time, which is indeed is 
not expected in a traditional Kolmogorov cascade. However, the MRI
generates strong, long-lived, extended azimuthal field bundles with relatively sharp gradients
 that do satisfy this condition \citep[][Paper~II]{Sano2007,McNally12a}.

\begin{table}[b!]
\caption{Run parameters \label{runpars}}
\centerline{\begin{tabular}{ccccc}
\hline \hline
$T_0$ (K)  & $\ell_0$ ($\mathrm{cm}$) & $t_{\eta}$(s) & $B_0'$ (Gauss) & $\beta_0$ \\
\hline
\multirow{4}{*}{$500$} & \multirow{4}{*}{$5\times 10^{20}$} & \multirow{4}{*}{$6 \times 10^{14}$} &
               $3$    &  $49.5$ \\
&   &   & $3.4$&  $38.5$ \\
&   &   & $4$    &  $27.8$ \\                                                             
&   &   & $5$    &  $17.8$ \\
\hline
\multirow{7}{*}{$990$} &  \multirow{7}{*}{$5\times 10^{9}$} & \multirow{7}{*}{$5 \times 10^3$} &
              $5$    &  $35.3$\\
&   &   & $5.5$ &$29.1$\\
&   &   & $6$    &$24.5$ \\
&   &   & $7.1$ &$17.5$\\
&   &   & $10$    &$8.8$ \\
&   &   & $15$    &$3.9$ \\
&   &   & $20$    & $2.2$ \\
\hline
\multirow{4}{*}{$1500$}&  \multirow{4}{*}{$2\times 10^{6}$} & \multirow{4}{*}{$5$} &
               $7.5$   &  $23.7$ \\
&   &   &  $12.5$ &  $8.5$  \\
&   &   &  $15$    &  $5.9$  \\                                  
&   &   &  $30$    &  $1.5$  \\
\hline
\multirow{4}{*}{$2000$}&  \multirow{4}{*}{$2.5\times 10^{5}$} & \multirow{4}{*}{$6$} &
               $12.5$&  $11.4$ \\
&   &  &  $19$ &  $4.9$  \\
&   &  &  $25$  &  $2.8$  \\                                                                 
&   &  &  $50$   &  $0.7$  \\
 \hline \hline
\end{tabular}}
\end{table}

We set the
background temperature $T_0$, from which we derive the boundary gas
pressure $P_{g,0}=\rho_0 k_B T_0/\mu m_p$, and total pressure
$P_0=P_{g,0}+B_0^2/(8\pi)$.  
We then derive the pressure, density and
temperature profiles from the magnetic field profile given in Equation~(\ref{initialB})
\begin{align}
P_g(x) &= P_0-B(x)^2/(8\pi) \\
\rho(x) &= \rho_0 \left(P_g(x)/P_{g,0}\right)^{1/\gamma}\\
T(x) &= T_0 \left(P_g(x)/P_{g,0}\right)^{1-1/\gamma} 
\end{align}
This initial condition includes a resistivity minimum at the origin due to the increased temperature there.
Note that \Eq{ionfrac} is a decreasing function of the density.  If we were to use an
isothermal hydrostatic initial condition, there would be an initial resistivity increase at the origin, which
can split the unstable region in
two, forming a swallowtail in a space-time diagram.  In that regard,
our adiabatic-hydrostatic initial  condition is also gentler than a
constant density-hydrostatic one due to the smaller spatial variation in the initial temperature.

While in the analysis of \Sec{linearanalysis} we assume a time-constant background temperature,
the spatial variation of $\eta$ may not dominate over the resistive diffusion of magnetic field in
\Eq{induction2}, especially at early time.  This results in a decaying current density at the origin, until
the instability has time to kick in.  While we could initialize our system with an inflow to confine the magnetic
field, this would cause significant compressive heating.

We make use of the symmetry in the problem along
the mid-plane of the current sheet in order to solve only one half of
the reconnection region.  We use zero-gradient boundary conditions on
the outer boundary (while pushing them towards infinity) and symmetric/antisymmetric boundary conditions as
appropriate at the origin. 
 \begin{figure*}
\begin{center}
\includegraphics[width=\linewidth]{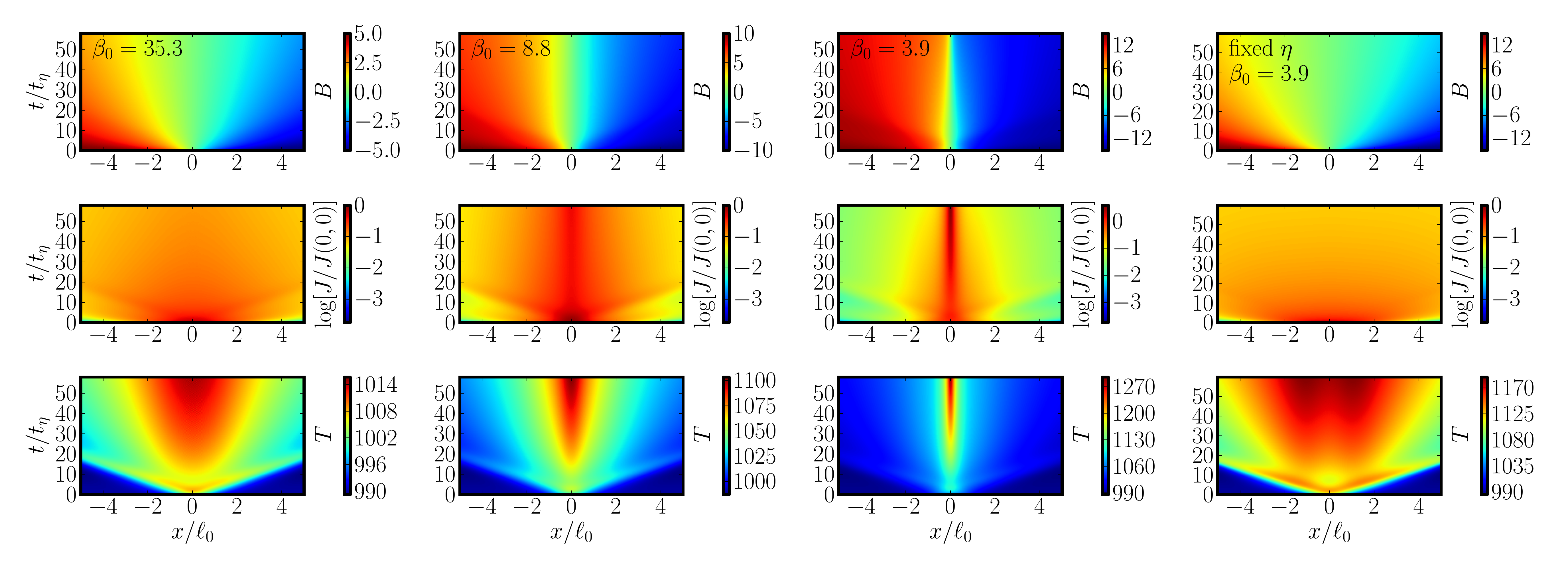}
\caption{Space-time evolution of $B$ in Gauss, self-normalized current density,
 and $T$ in K for three different values of initial $\beta_0$,
with a background temperature of $T_0=990$~K. 
In the fixed $\eta$ case, 
the ionization level is set to that from the background non-thermal
ionization.
\label{BJT1k}}
\end{center}
\end{figure*}
 
\section{Results}
\label{sec:results}

In \Fig{BJT1k} we show the space-time evolution of the instability as
a function of the ratio of thermal to magnetic pressure
\EQ
 \beta=8\pi \rho_0 k T_0/(\mu m_p B_0^2),
\EN
with and without temperature
dependent resistivity.  In the third column
we see the typical nature of the instability: a strong
current sheet develops, shown by the narrowing of the magnetic field.  When the 
temperature dependent ionization is turned off (column 4)
the reconnection region diffuses outwards normally.
We also see a difference between this system and the idealized one of \Sec{linearanalysis}:
the current density in the center spreads resistively at early times so the background current
is not constant in time.
The difference between columns 3 and 4
 shows the importance of
treating the temperature dependence of the resistivity in this system. 

\subsection{Growth Rate}
\label{sec:growth}

\begin{figure}
\centering\includegraphics[width=\linewidth]{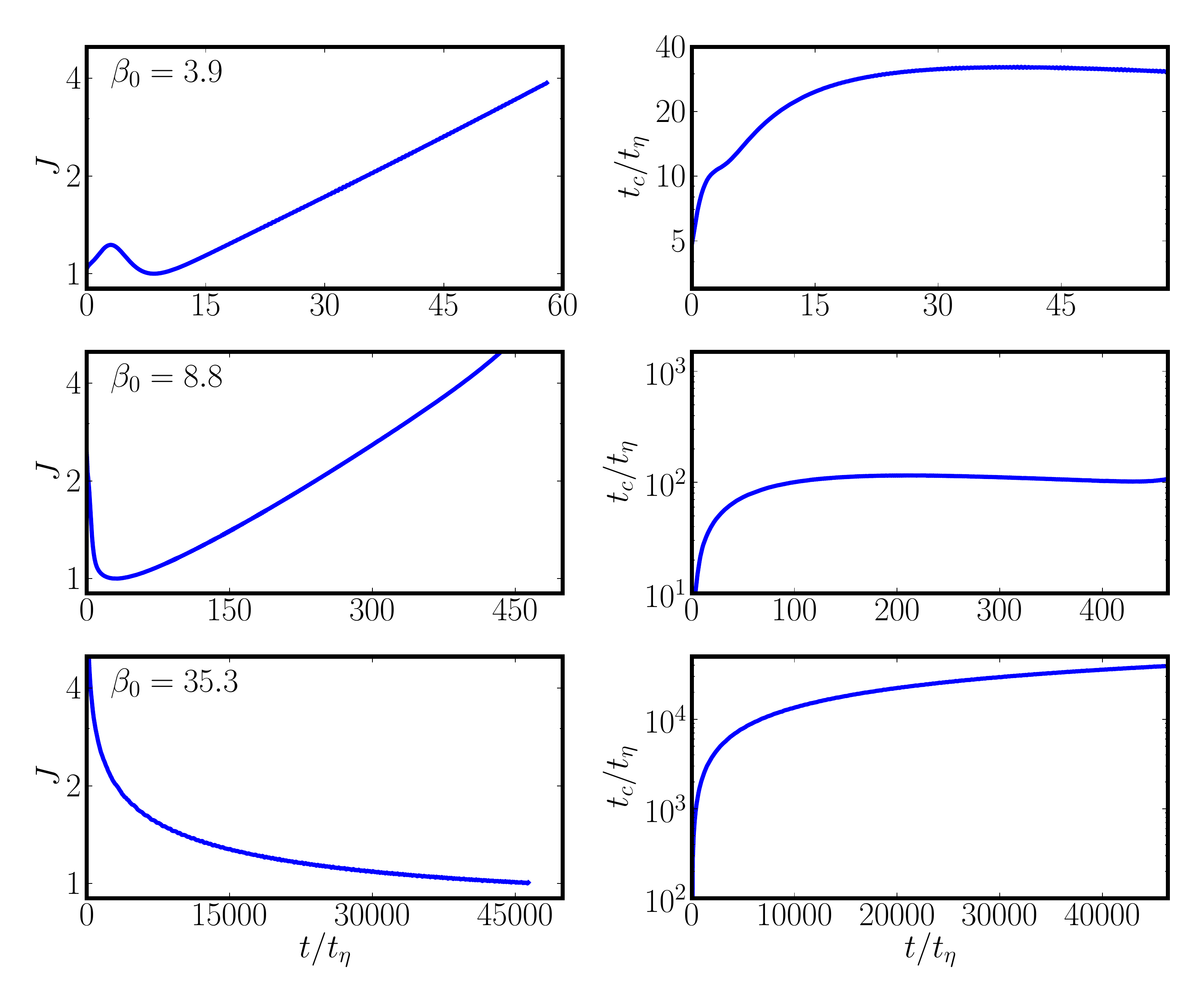}
\caption{
Top panels: $\beta_0=4.9$; Middle panels: $\beta_0=18.6$; Bottom panels: $\beta_0=37.5$.
Left panels: Time variation of J at the origin, normalized to its minimum value.
Right panels: the dynamically defined timescale
$t_c$, normalized to the resistive time $t_{\eta}$.  All
for three values of $\beta_0$  given in the panels and
a background temperature of $T_0=990$~K.
\label{timenorm}}
\end{figure}
We can use
the linear growth rate given by \Eq{instabgrowth} to estimate
the growth rate of the instability in the nonlinear regime
reached in the simulations.  In this regime, the
value of  the growth rate 
$\lambda(x,t)$ depends strongly on both position and time because of the
variation of the parameters, especially $\eta$,
but also the initial resistive spreading of the current density (see \Fig{BJT1k}).
We extend to the nonlinear case 
by computing the value of $\lambda(0,t)$ at the center of the
current sheet and examining it to see if it saturates to a
constant value in the nonlinear regime that provides a reliable
estimate of the actual growth rate.  We define a timescale
$t_c(t)=\lambda^{-1}(0,t)$ that we use to test this hypothesis.

In the following computation of $t_c$, we use 
$\ell \equiv \max(B)/J(x=0)$ as the approximate, time-varying,
width of the current sheet, taking the place of $k^{-1}$ in the linear
analysis, and
\begin{align}
&\beta_c=\frac{8 \pi n_n k_B T}{\max(B^2)}, \\
&t_{r,c}= \ell^2/\eta, \\
&t_{h,c}=\frac{\beta_c}{\gamma-1}\frac{t_{r,c}}{2}, \\
&t_{\eta',c}=\frac{T}{T_i} t_{h,c},
\end{align}
where all spatially varying quantities are determined at $x=0$.
Solving \Eq{instabgrowth} for $\lambda$ using the above definitions,
we find 
\begin{align}
(1/t_c) =&\frac{1}{2t_{r,c}}\left\{\vphantom{\left[\left(\frac{2T_i}{T}\right)^2\right]^{1/2}}
-\left(1+\frac{2T_i}{T}\frac{\gamma-1}{\beta_c}\right)+ \right. \label{instabgrowth2}  \\ 
&\left. \left[1+\frac{12T_i}{T}\frac{\gamma-1}{\beta_c}+
    \left(\frac{2T_i}{T}\frac{\gamma-1}{\beta_c}\right)^2\right]^{1/2} \right\}^{-1}.  \nonumber
\end{align}

In \Fig{timenorm} we plot $J(0)$, and $t_c$, normalizing by the resistive time at the start
of the simulation,
$t_{\eta} \equiv t_{r,c}(t=0)$.
We can see that $t_c(t)$ is reasonably well behaved and indeed
has a defined plateau that occurs after the onset of current density growth. 
On the other hand, it initially grows significantly because of
 the resistivity drop caused by Ohmic heating.  As $t_c$ 
possesses a defined maximum value in systems that show current density growth,
we will use its maximum value 
 for our estimate of the growth rate.
Unfortunately, this is not strictly well defined in cases that do not
show current density growth (e.g. \Fig{timenorm}, bottom panel). 

\begin{figure}
\begin{center}
\includegraphics[width=\columnwidth]{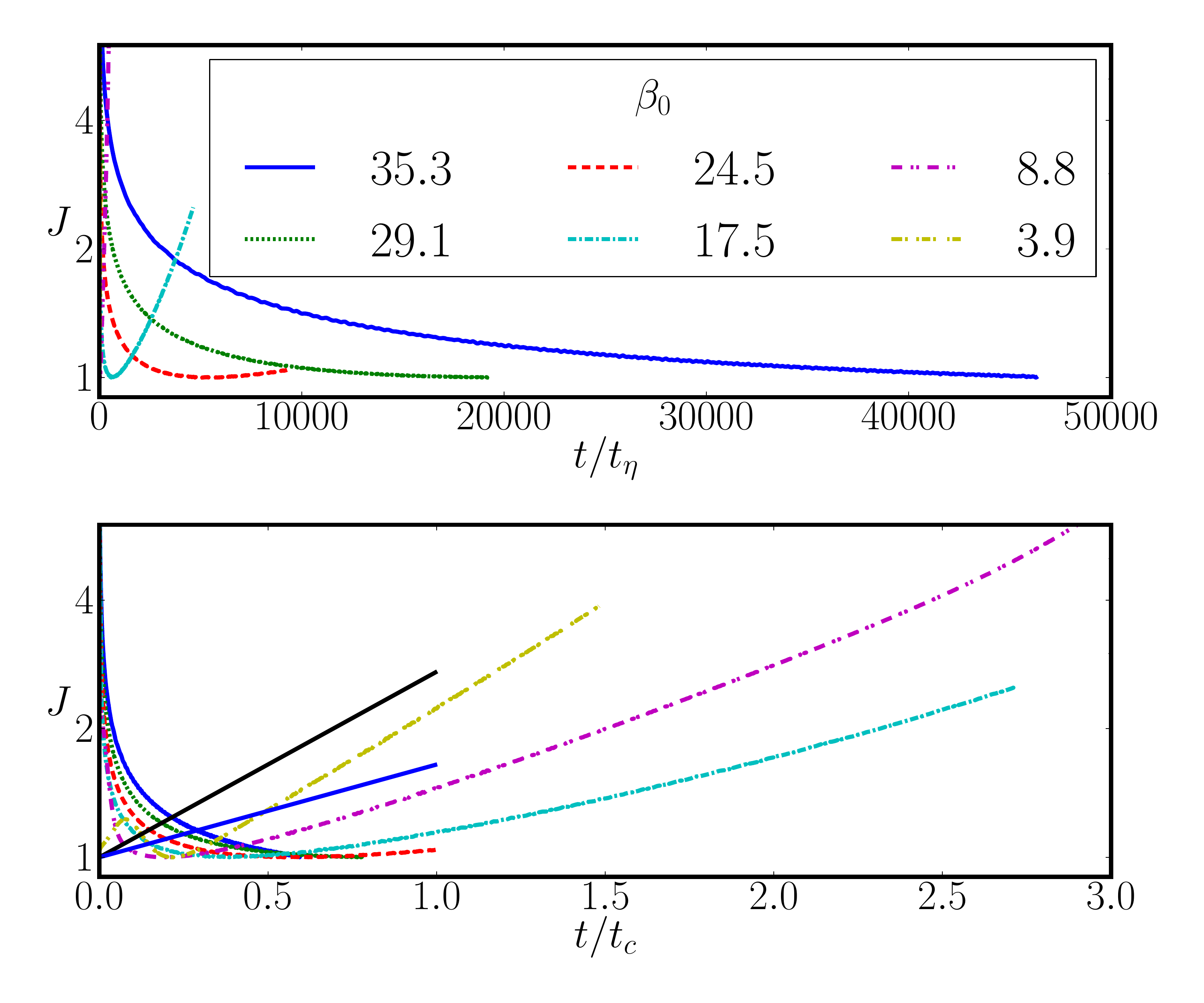}
\caption{Time series of J at the origin, normalized to its minimum
  value for two different time normalizations and six values of
  $\beta_0$. 
The background temperature of these models is $T_0=990$~K.  The solid,
straight curves show $\exp(t/t_c)$ and $\exp(t/2t_c)$.}
\label{1k}
\end{center}
\end{figure}

\begin{figure}
\begin{center}
\includegraphics[width=\columnwidth]{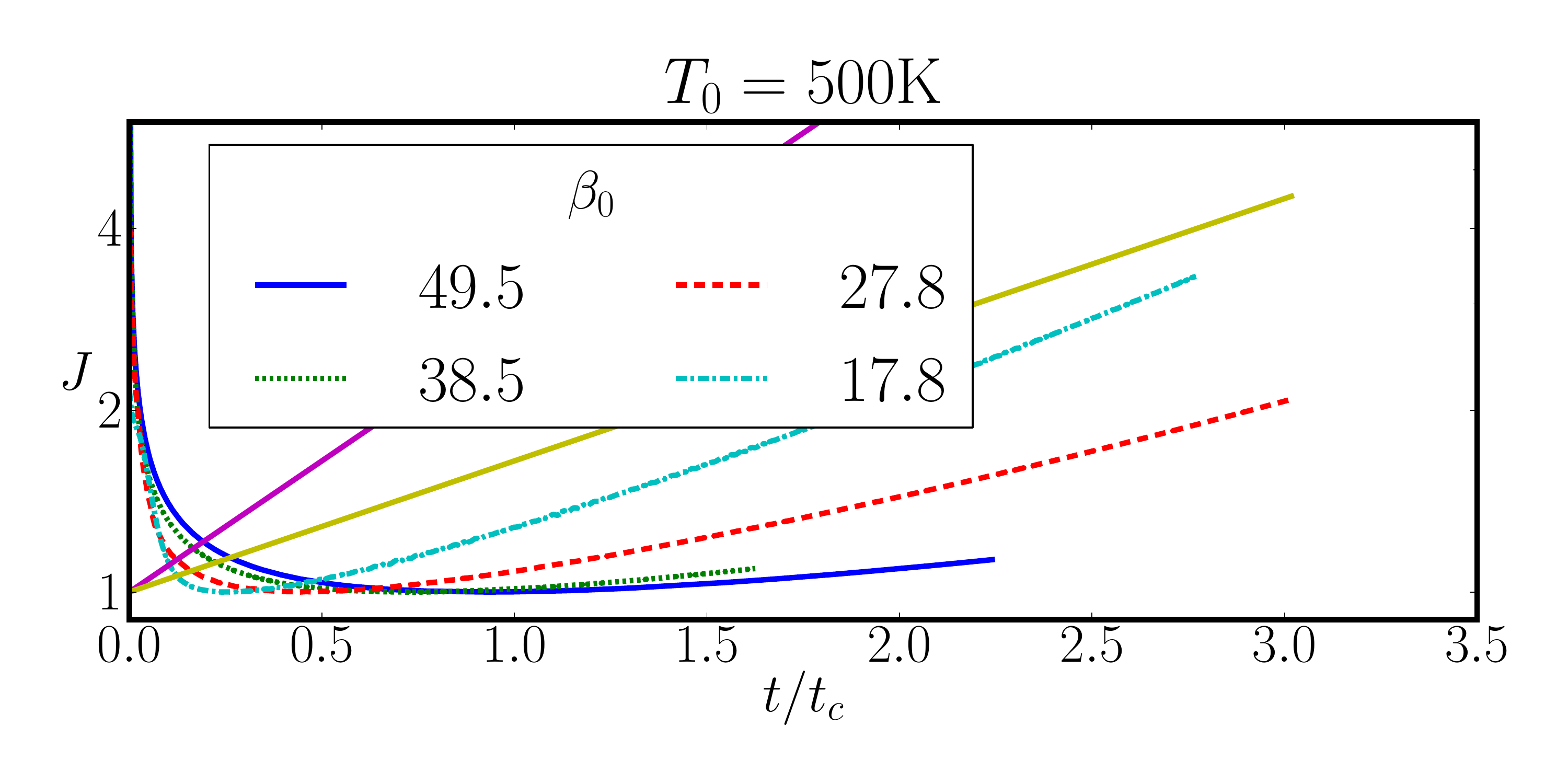}
\includegraphics[width=\columnwidth]{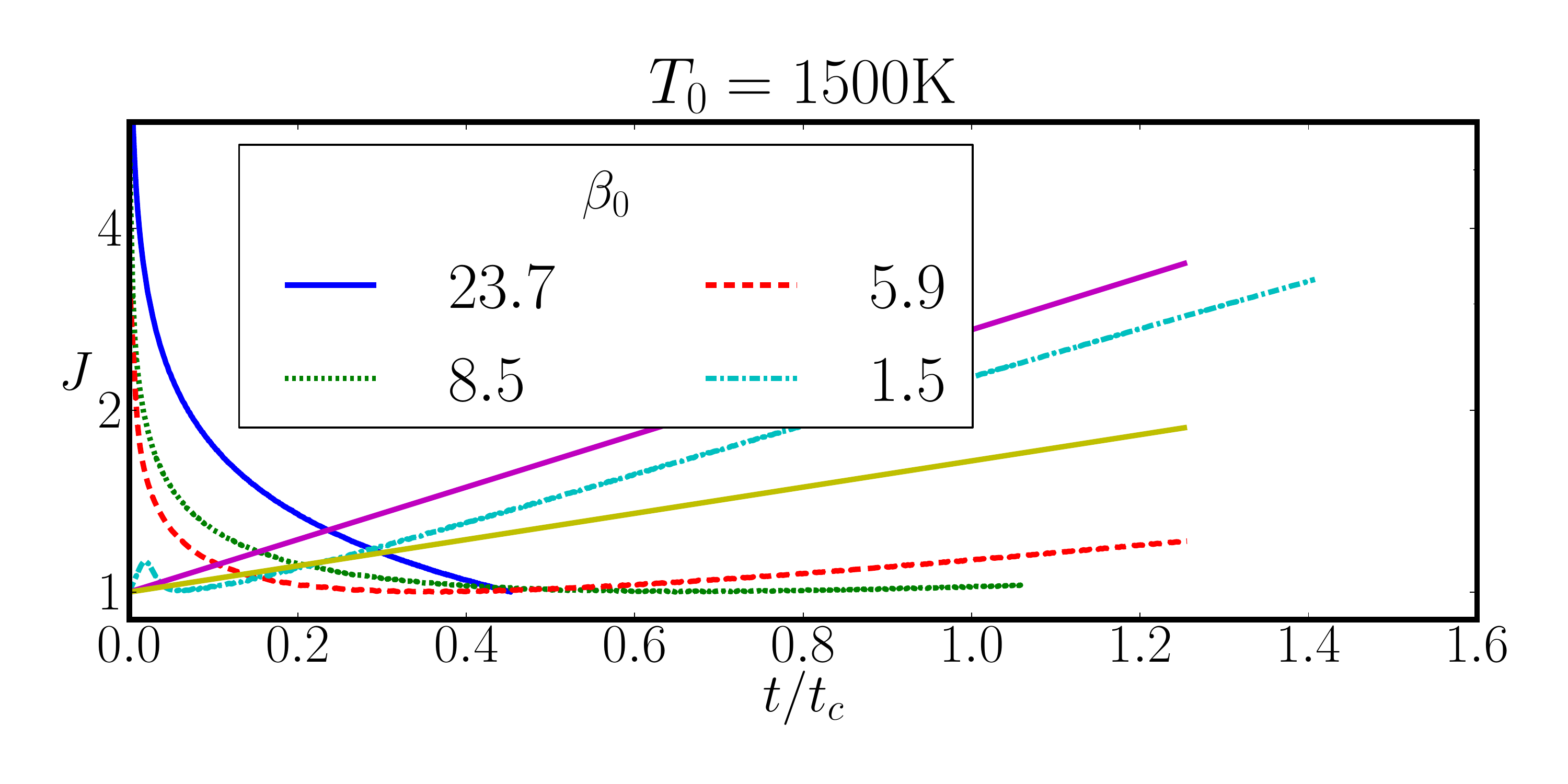}
\includegraphics[width=\columnwidth]{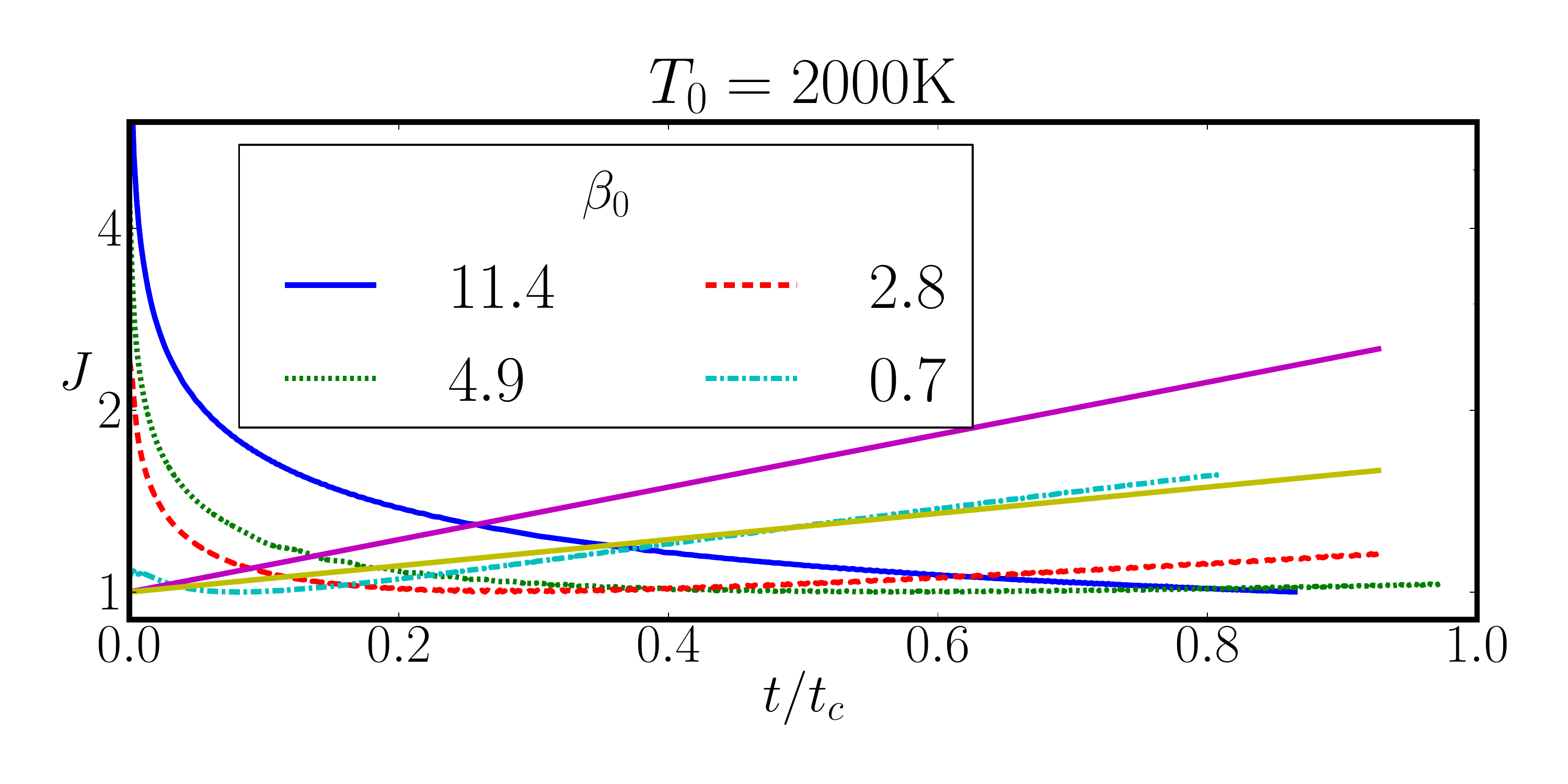}
\caption{Time series of J at the origin, normalized to its minimum
 value, for three background temperatures, $T_0=500,\, 1500$, and 2000~K.}
\label{multitemp}
\end{center}
\end{figure}

In \Figs{1k}{multitemp} we show the evolution of the central
current density $J$ for four temperatures and varying $\beta_0$.  Further, 
we include the functions $\exp(t/t_c)$ and $\exp(t/2t_c)$.  If the analysis of \Sec{linearanalysis}
were exactly applicable with our definition of $t_c$, 
then the growing instabilities would have the slope of the former.
It is clear that,
although $t_c$ is a reasonable estimate of the timescale for
instability growth,
the curves of growing $J$ do not all have the same slope, and
$t_c$ overestimates growth rates in some cases. 
Recall that $t_c$
is not well defined for the runs
that fail to go unstable. 

The fact that $t_c$ is a good estimate of the growth rate of the
instability is perhaps
surprising in light of 
 the behavior of the resistivity (\Fig{rhoplot}, bottom panel).
 As the instability grows
the resistivity at the origin drops by nearly two orders of magnitude,
while the current sheet gets strongly concentrated: the system 
 has become
strongly nonlinear.  A possible explanation 
for the continued relevance of the linear analysis, however,
is that the symmetry of the model problem maintains the
validity of the assumptions in \Sec{linearanalysis}.  In
particular, 
the current sheet concentrated by the instability has a flat current density
at the origin that acts as the background current density for further growth.       
We discuss this further in \Sec{saturation} below.

\subsection{Stability Criterion}

From \Sec{linearanalysis} we might expect that all our simulations should be
unstable due to the lack of cooling.  While \Figs{1k}{multitemp}
show runs that have not gone unstable, it is unclear whether this is
due to the additional physics that we have added to the problem
changing the condition for instability,
or merely inadequate run time (note that longer run times require
larger boxes to exile the boundaries to infinity).
The growth rate does appear to drop with increasing $\beta_0$
 even when
normalized to $t_c$.
This slower growth may be due to the lower value of $t_{r,c}/t_c$ in the
high $\beta_0$ case (see Equation~\ref{instabgrowth2}).  Unlike in the linear analysis,
the resistive time also acts to spread the background current sheet and will
decrease the instability's growth rate, and may halt it altogether.

However, in
\Fig{1k}, top panel, the curve associated with $\beta_0=9.9$ is just
distinguishable from the left axis, while the curve associated with
$\beta=4.9$ is not.  Clearly the 
 increase in instability growth time with $\beta_0$
is pronounced, with the lowest obviously unstable $\beta_0=25.6$
presented having its instability kick in at $t\sim 5000t_{\eta}$.
We expect that
higher $\beta_0$s, if unstable, would require even longer.  Even if
the higher $\beta_0$ runs are eventually unstable, it is unlikely to
be a matter of practical concern in a physical system.
We attribute this to the longer interval in which the resistive spreading term in \Eq{induction2} dominates over the
instability term, resulting in an ever increasing $\ell$, and so an ever
decreasing $\lambda$.

\subsection{Saturation}
\label{saturation}

\begin{figure}
\begin{center}
\includegraphics[width=\columnwidth]{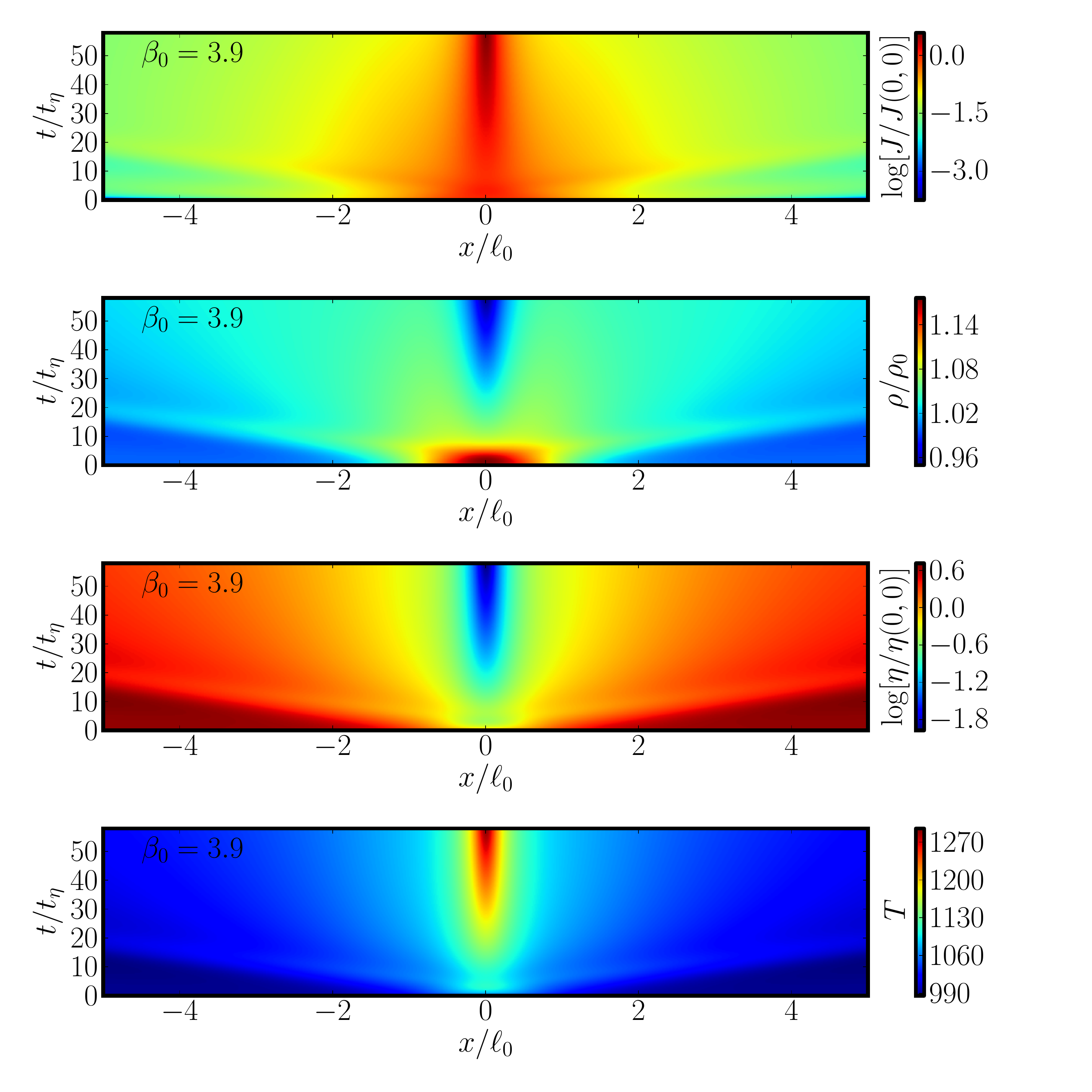}
\caption{Self normalized current density, mass density, and resistivity, and
temperature in Kelvins,
for $\beta_0=3.9$ and a background temperature of $990$~K.  Note that 
the density drops in the central region implying mass 
outflow from the central region even while magnetic flux is transported 
inwards 
(as traced by the growing current).
\label{rhoplot}}
\end{center}
\end{figure}

\begin{figure}
\begin{center}
\includegraphics[width=\columnwidth]{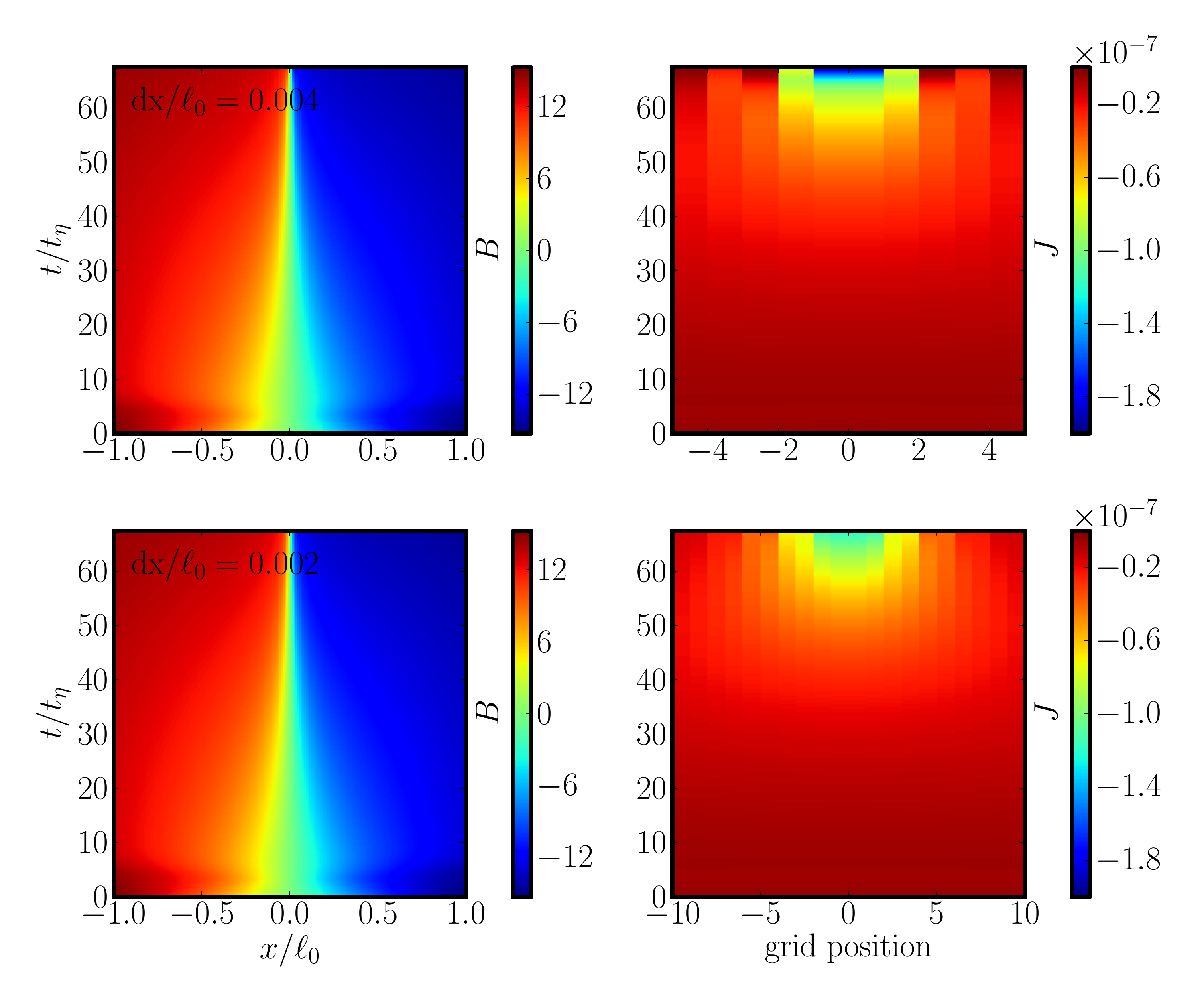}
\caption{Magnetic field (Gauss), and current density plots
for $\beta_0=2.2$ and a background temperature of $990$~K.  Top panels: $dx/\ell_0=0.004$; Bottom panels: $dx/\ell_0=0.002$.
On the right panels, the x-axes are in grid points rather than position (logarithmic spacing) but the plots cover the same physical extent.
We can see the high resolution run continuing to narrow after the low
resolution run has hit the grid scale and started ringing,
as shown by the sooty dashes at the very top of the top-right panel. 
\label{satplot}}
\end{center}
\end{figure}

In the absence of cooling and ever decreasing resistivity, 
it is not clear how the instability can saturate.
So long as the current density $J$ remains differentiable at the
origin, symmetry requires that $\parrx J=0$, so 
the approximation assumed in \Sec{linearanalysis} remains good.
The 
saturated instability must
maintain the constant background current assumed.  Further, as we can see in \Fig{rhoplot}, while
the system is compressible, it is not very compressible, with $\rho T$ approximately constant.
This sustained satisfaction of the linear instability criterion may explain why
the time-scale estimate $t_c$, which is determined from a linear stability
analysis, performs as well as it does despite the non-perturbative evolution of the system
shown in \Fig{rhoplot}.

In \Fig{satplot} we show
the evolution of the same system at two different resolutions,
with the magnetic field plotted according to
physical position on the left, and on the right, the current density
plotted according to 
 position on the logarithmically spaced grid.
The upper panels show models with minimum grid resolution
$dx=0.004\, \ell_0$ while the lower panels have $dx=0.002\, \ell_0$.
The physical extent
associated with the right panels is the same for the two resolutions
and the same mapping of current density to color is used.  
In the left panels, we show the initial field, and the narrowing
associated with the growing current sheet, 
which evolves similarly at
both resolutions.  In the right panels, we show the current density
reaching the innermost grid point (the left side of the plot is the
mirror image of the right) at low resolution, and associated ringing,
while the high resolution run continues to narrow.
It too will eventually narrow to below its grid resolution and be
subject to the same numerical instability.

In the absence of cooling, or changing physics, such as an alteration
to the resistivity equation, it appears that the instability will not
saturate.
If the background current density is constant in time and
differentiable at the origin, symmetry requires that it
have a constant component which is linearly unstable
in the absence of cooling.
Formally, the $\eta \nab^2 \BB$ term in the induction equation~(\ref{induction})
prohibits non-differentiable current densities,
so that the actual saturation mechanism must involve changing physics.

One clear possibility is that the action
of cooling in combination with a change in the temperature dependence of $\eta$
should halt the instability, as cooling will become
stronger at higher temperatures while the resistivity temperature
gradient weakens.  At 
high temperatures, it is expected that
the temperature dependence 
of resistivity will change from that given by
Equations~(\ref{spitzerresistivity}) and (\ref{ionfrac}).  As the
ionization fraction approaches full ionization, the temperature
dependence of the resistivity will weaken.  If the ionization fraction
saturates, then the current sheet instability itself will saturate.

\section{Discussion and Conclusions}
\label{sec:discuss}

We have shown that in slab-symmetric reconnection, a strong  inverse
temperature dependence of the resistivity can lead to an instability that
concentrates the current in a narrow, high temperature, low
resistivity sheet.  This scenario is the polar opposite of the more
common situation where the resistivity appears to increase
inside current sheets through some anomalous resistivity
\citep{KrallLiewer71,SatoHayashi79}.  Unlike many reconnection
scenarios, the inward transport of the magnetic field  in our case 
is resistive rather than advective in nature, sidestepping issues of fluid pile up
that occur with advective field transport, as demonstrated by
\Fig{rhoplot}, where the central density declines with time.

However, rather than speeding the dissipation of magnetic energy into
heat, in one dimension the total dissipation rate 
actually falls,
thanks to the formation of small volumes with very low resistivity,
even though heating increases sharply within those regions.
This raises interesting questions about
the structure of magnetic turbulence in three-dimensional 
systems with similar temperature dependent resistivities.

The instability does not grow extremely fast as can be seen in \Fig{timenorm},
which shows that the instability growth rate estimate $t_c$ is significantly slower than
the background resistive broadening time of the current sheet $t_{\eta}$. Nevertheless,
as seen clearly in the third column of \Fig{BJT1k}, the instability can grow on timescales
of tens of resistive times. 
For this to occur, the strength of the magnetic field must allow
rapid heating, and external large-scale dynamics must not tear
the current sheet apart.  Both of these conditions appear
reasonable for disks subject to the MRI \citep{Sano2007}.
Further, it is clear from \Fig{BJT1k} that a fully self-consistent 
analysis is needed for any MRI active region in a protoplanetary
disk whose magnetic field and temperature flirt with $\beta \sim 1-4$ and $T \sim 1000$~K.
We have only considered growth of the instability
 in approximations to current sheets that the MRI generates
in the absence of our instability. 
Such a self-consistent approach
will be difficult considering the large range in dissipation parameters that must be resolved
and the large spatial scale separation between the turbulence (larger than $\ell_0$) and
the narrowed current sheets (much smaller than $\ell_0$, Fig \ref{satplot}).

We expect such self-consistent systems
to show the concentration of current into localized regions with
high current and temperature (either
two-dimensional sheets or one-dimensional tubes), with the rest of space taken up by
almost force-free magnetic fields.
As the concentrated current regions have low resistivity,
they could potentially have long lifetimes, 
 perhaps much 
longer than that associated with the high
wavenumber tail of a subsonic turbulent cascade.  Where the magnetic
field energy does not exceed equipartition with
the fluid kinetic energy, the bending of the magnetic field will create new
current structures that fence in the magnetic field configuration, with the potential
for extremely large and localized Lorentz forces because of the highly
concentrated current densities. 

While we have shown that, in sufficiently restricted circumstances,
this instability occurs for any inverse relationship between the
resistivity and the temperature, we have also shown that it can take a
prohibitive time to set in.  In practice it appears that the growth
rate estimate of \Eq{instabgrowth}, while
   sometimes
an overestimate, is accurate
within factors of a few.  Unfortunately, evaluating it requires that
the instability be growing, and any initial transients can result in
strong overestimates of the growth rate (see \Fig{timenorm}, early
times).

We explore the peak temperatures achieved by this instability further
in Paper II, in particular by including
radiative transfer and a fuller
treatment of thermal ionization.  However
the significance for protoplanetary
temperature structures in the inner MRI active disk is
   already clear from the work presented here.

Although we 
have considered this effect from the perspective of protoplanetary
disks, it should occur in any system with an adequately strongly
increasing conductivity dependence on temperature when compared to
available cooling.  Candidates include cool stellar surfaces with
poorly ionized hydrogen, and even planetary atmospheres, as recently
suggested by \citet{Menou12}.  

\acknowledgements We thank D. Ebel, W. Lyra, and J. Oishi for useful
discussions.  A.H. was partly supported by a Kalbfleisch Fellowship
from the American Museum of Natural History. We acknowledge support
from the NSF through CDI grant AST08-35734 and AAG grant AST10-09802.


\end{document}